\documentclass[twocolumn,showpacs]{revtex4}
\usepackage{times,xspace}
\usepackage{amsbsy,amssymb,amsmath,bm}
\usepackage{graphicx,color,epsfig,rotate}
\usepackage{fancyhdr}

\def\bbbc{{\mathchoice {\setbox0=\hbox{$\displaystyle\rm C$}\hbox{\hbox
to0pt{\kern0.4\wd0\vrule height0.9\ht0\hss}\box0}}
{\setbox0=\hbox{$\textstyle\rm C$}\hbox{\hbox
to0pt{\kern0.4\wd0\vrule height0.9\ht0\hss}\box0}}
{\setbox0=\hbox{$\scriptstyle\rm C$}\hbox{\hbox
to0pt{\kern0.4\wd0\vrule height0.9\ht0\hss}\box0}}
{\setbox0=\hbox{$\scriptscriptstyle\rm C$}\hbox{\hbox
to0pt{\kern0.4\wd0\vrule height0.9\ht0\hss}\box0}}}}

\newcommand{\diag}{\mbox{dg}}

\newcommand{\ignore}[1]{}
\newcommand{\mComment}[1]{}
\newcommand{\gComment}[1]{}
\newcommand{\jComment}[1]{}
\newcommand{\rComment}[1]{}
\newcommand{\lComment}[1]{}

\renewcommand{\mComment}[1]{\textcolor{blue}{Manny: #1}}
\renewcommand{\gComment}[1]{\textcolor{red}{Gerardo: #1}}
\renewcommand{\jComment}[1]{\textcolor{green}{Jim: #1}}
\renewcommand{\rComment}[1]{\textcolor{magenta}{Ray: #1}}
\renewcommand{\lComment}[1]{\textcolor{purple}{Rolando: #1}}

\begin{document}

\title{Generalized Elitzur's Theorem and Dimensional Reductions}
\author{C. D. Batista and Zohar Nussinov}
\address{Theoretical Division,
Los Alamos National Laboratory, Los Alamos, NM 87545}

\date{Received \today }

\begin{abstract}
We extend Elitzur's theorem to systems with symmetries 
intermediate between global and local.
In general, our theorem formalizes the idea of {\it dimensional reduction}.
We apply the results of this generalization to many systems that are
of current interest. These include liquid crystalline phases 
of Quantum Hall systems,
orbital systems, geometrically frustrated spin lattices,  
Bose metals, and models of superconducting arrays. 

\end{abstract}


\maketitle

\section{Introduction} 

Symmetry occupies a central and indispensable role in physics. 
Amongst others, it is the key concept behind the theory of phase transitions 
and critical phenomena. Most of the known phase transitions 
break one or more symmetries. When symmetry breaking
occurs, the mean value of a quantity which is not invariant 
under the symmetry group of the theory becomes 
``spontaneously'' non-zero in the broken symmetry phase.
Such a quantity is called the {\it order parameter}. As well 
appreciated, the dimension $D$ of the system and the symmetries 
of the order parameter characterize different universality classes
and play a pivotal role in determining 
whether a particular phase transition may or may not exist. 
In particular, zero or one dimensional 
theories with short range interactions
cannot exhibit a phase transition at any finite 
temperature. Additionally, the Mermin-Wagner theorem 
\cite{Mermin} states that a continuous symmetry cannot be spontaneously 
broken at any finite temperature for two-dimensional 
theories with finite range interactions. On the other hand, Elitzur 
\cite{Elitzur} demonstrated that a spontaneous breaking of
a {\it local} symmetry is not possible. Below, we will 
show that Elitzur's theorem is a consequence of a reduction to zero 
of the effective dimension of the gauge invariant 
theory. Moreover, we will show that from the point of view of the 
non-invariant 
gauge fields, the presence of a ``$d$-dimensional gauge symmetry''
(see definition below) reduces the effective dimension of the 
theory from $D$ to $d$.
The dimension $d$ is intermediate between local gauge systems ($d=0$) 
and global 
gauge symmetries ($d=D$).

How is it possible that symmetry not only determines the nature 
of the transition but 
also its effective dimension? The answer to this question requires 
a definition of what  
we mean by {\it local} and ``$d$-dimensional gauge symmetries''. 
A local gauge symmetry has its roots in
a group of operations
which leave the theory invariant 
and in which the minimum 
non-empty set of fields influenced by any of these operations 
occupies a finite spatial region.  
A ``$d$-dimensional gauge symmetry''
is any symmetry operation which leaves the Hamiltonian
(or action) invariant such that the minimum non-empty 
set of fields which are changed 
under the symmetry operation occupies a $d$-dimensional region. 
In accord with this definition, a local 
gauge symmetry corresponds to the particular case $d=0$. 
The above question is motivated 
by the recent interest in a considerable number of theories 
that are invariant under a 
$d$-dimensional gauge group with dimension $d \geq 1$ 
\cite{Harris,Xu03,Xu04,Zohar,Mishra,NBCv,BCN,NBCF}. As was already noticed 
\cite{Harris,Xu03,Xu04}, some aspects of these theories indicate that the
dimension is effectively reduced. In this paper we formalize this 
notion by proving a generalized
Elitzur's theorem for any value of $d$.

We will prove that the mean value of a local quantity which is non-invariant 
under a $d$-dimensional gauge group can be bounded by another mean 
value of the 
same quantity which is computed for an effective $d$-dimensional theory. The 
gauge transformations become a global symmetry of the effective 
theory and the 
range of the interactions in the original theory is preserved. This proof 
formalizes the idea of ``dimensional reduction'' and gives rise to four 
important 
corollaries. The first corollary is Elitzur's theorem. The second 
states that 
a $d=1$ gauge symmetry cannot be broken at any finite temperature if the the 
theory only includes finite range interactions. The third corollary is 
a natural 
consequence of the Mermin-Wagner theorem: 
a {\it continuous} $d=2$ gauge symmetry 
cannot be broken at any finite temperature if the interactions 
are of finite range. 
The fourth corollary extends Elitzur's theorem to certain systems in which 
at least an {\em emergent} low energy gauge-like symmetry appears. By emergent symmetry
we mean a group transformations which become a symmetry of the Hamiltonian only within its lowest-energy 
subspace \cite{BO}.

Recently, theories containing $d$-dimensional gauge symmetries 
with $d \geq 1$ became the focus 
of attention in different areas of physics. The 
Kugel-Khomskii model that describes the interplay between spin and 
orbital degrees of freedom 
in transition metals as well as orbital only variants
embodying the Jahn-Teller interactions 
often display $d=1,2$ gauge-like symmetries 
\cite{Zohar,Mishra,NBCv,BCN,NBCF}.
The well known planar orbital compass model 
\cite{Zohar, Mishra,NBCF} is invariant under a $d=1$ $Z_2$-gauge group. 
The two-dimensional frustrated 
Ising magnet studied by Xu and Moore \cite{Xu03,Xu04} 
provides another example of a $d=1$ $Z_2$-gauge 
invariant theory. There are also examples of frustrated magnets 
\cite{BT} for which the $d=1$ $Z_2$-gauge 
group is an {\it emergent symmetry} \cite{BO}. 
Here, the dimensional reduction may
occur at low energies. Similar symmetries appear in 
ring exchange Bose metals \cite{Arun02},  
Quantum Hall liquid crystalline phases
\cite{lawler}, \cite{rad}, lipid 
bilayers with intercalated DNA strands \cite{ohern},
and "sliding Luttinger liquids" 
\cite{emery2000}.

The physical roots of our theorem and its corollaries are
precisely the same  fluctuations which generically inhibit 
order in low dimensions. Just as domain wall
entropy eradicates any viable order
in one dimensional systems having short range
interactions, domain walls on different
sliding phases of many of the above examples
inhibit discrete d=1 gauge symmetry breaking at
any finite temperature.Similar considerations
apply to continuous d= 1,2 dimensional gauge
symmetry systems whose order is destroyed by
low energy excitations.

\section{A Generalization of Elitzur's Theorem} 

We will start by considering the case of classical theories. 
As we will see later, the generalization to 
the quantum case is very simple. Before enunciating our theorem 
it is convenient to give a more precise formal definition 
of a $d$-dimensional gauge group. From now on we will 
consider theories which are defined on a lattice ${\Lambda}$. 
The extension to the continuum case is straightforward.  A 
$d$-dimensional gauge symmetry of a theory 
characterized by a Hamiltonian $H$ is  a group of symmetry 
transformations of $H$ such that the minimal non-empty 
set of fields $\phi_{\bf i}$ that is changed under the group 
operations occupies a $d$-dimensional subset 
${\cal C} \subset \Lambda$. The index ${\bf i}$ denotes the sites 
of the lattice ${\Lambda}$. For instance, 
if a spin theory is invariant under flipping each individual spin the 
corresponding gauge symmetry is zero-dimensional 
or local. Of course flipping a chain of spins is also a symmetry, 
but the chain is not the minimal non-trivial subset 
of spins that can be flipped. In general, 
these transformations can be expressed as:
\begin{equation}
{\bf U}_{lk} = \prod_{{\bf i} \in {\cal C}_l} {\bf g}_{{\bf i}k},
\label{tran}
\end{equation}
where ${\cal C}_l$ denotes the subregion $l$, ${\cal C}_l \subset \Lambda$,
and  
${\Lambda}= \bigcup_{l} {\cal C}_l$.

{\it Theorem: The absolute mean value of any local quantity 
(i.e. involving only a finite number of fields)
which is not invariant under a $d$-dimensional gauge symmetry group 
$G$ of the $D$-dimensional Hamiltonian $H$
is equal or smaller than the absolute mean value of the same quantity 
computed for a $d$-dimensional
Hamiltonian ${\bar H}$ which is globally invariant under the group $G$ 
and preserves the range 
of the interactions in $H$}. Non invariant here means that the 
quantity under 
consideration, $f(\phi_{\bf i})$ , has no invariant component:
\begin{equation}
\sum_k f[{\bf g}_{{\bf i}k}(\phi_{\bf i})] = 0.
\label{nonin}
\end{equation}
For a continuous group Eq.(\ref{nonin}) has to be replaced by
$\int f[{\bf g}_{\bf i}(\phi_{\bf i})] d{\bf g} = 0$.
To determine if there is a spontaneous symmetry
breaking in the thermodynamic limit, the mean value of
$f(\phi_i)$ has to be computed in the following way:
\begin{equation}
\langle f(\phi_{\bf i}) \rangle = lim_{h\rightarrow 0} 
lim_{N \rightarrow \infty} 
\langle f(\phi_{\bf i}) \rangle_{h,N},
\label{limit}
\end{equation}
where $\langle f(\phi_i) \rangle_{h,N}$ is the mean
value of $f(\phi_i)$ computed on finite lattice of 
$N$  sites and in the presence of a symmetry breaking field $h$. 
Since ${\Lambda}= \bigcup_{l} {\cal C}_l$, 
the site ${\bf i}$ belongs at least to one
set ${\cal C}_j$. It is convenient to rename the fields 
in the following way: 
$\phi_{\bf i}=\psi_{\bf i}$ if ${\bf i} \notin {\cal C}_j$ and 
$\phi_{\bf i}=\eta_{\bf i}$ if ${\bf i} \in {\cal C}_j$.The mean value 
$\langle f(\phi_{\bf i}) \rangle_{h,V}$ is given by:
\begin{eqnarray}
\langle f(\phi_{\bf i}) \rangle_{h,N} =
\frac{\sum_{\{ \phi_{\bf i} \}}  f(\phi_{\bf i}) e^{-\beta H(\phi)} 
e^{-\beta h \sum_{{\bf i}} \phi_{\bf i}}}
{\sum_{\{ \phi_{\bf i} \}} e^{-\beta H(\phi)-\beta h \sum_{{\bf i}} 
\phi_{\bf i}}}=
\nonumber \\
\frac{\sum_{\{ \psi_{\bf i} \}} 
z_{\{ \psi \}} e^{-\beta h \sum_{{\bf i} \notin {\cal C}_j} \psi_{\bf i}} 
[\frac {\sum_{\{ \eta_{\bf i} \} } f(\eta_{\bf i})  
e^{-\beta H(\phi)-\beta h \sum_{{\bf i} \in {\cal C}_j}  
\eta_{\bf i}}}{z_{{\{ \psi \}}}}]}
{\sum_{\{ \psi_{\bf i} \}} 
z_{{\{ \psi \}}} e^{-\beta h \sum_{{\bf i} 
\notin {\cal C}_j} \psi_{\bf i} }  }
\label{mast}
\end{eqnarray}
with,
\begin{equation}
z_{{\{ \psi \}}}= \sum_{\{ \eta_{\bf i} \}} 
e^{-\beta H(\psi,\eta)-\beta h \sum_{{\bf i} 
\in {\cal C}_j}  \eta_{\bf i}}.
\end{equation}
From Eq.(\ref{mast}):
\begin{equation}
|\langle f(\phi_{\bf i}) \rangle_{h,N}| \leq \mid   
\frac {\sum_{\{ \eta_{\bf i} \} } f(\eta_{\bf i})  
e^{-\beta H({\bar \psi},\eta)-\beta h \sum_{{\bf i} 
\in {\cal C}_j}  \eta_{\bf i}}}{z_{{\{ {\bar \psi} \}}}} \mid, 
\label{final}
\end{equation}
where $\{ {\bar \psi} \}$ is the particular configuration 
of fields ${\psi_{\bf i}}$ that maximizes the 
expression between brackets in Eq.(\ref{mast}). 
$H({\bar \psi},\eta)$ is a $d$-dimensional Hamiltonian for 
the field variables ${\eta}$ which is invariant under the 
{\it global} symmetry group $G_j$ of transformations
${\bf U}_{jk}$ over the field $\eta$. 
We can define ${\bar H}(\eta) \equiv H({\bar \psi},\eta)$. 
The range of the interactions between the 
$\eta$-fields in ${\bar H}(\eta)$ is clearly the same 
as the range of the interactions between the 
$\phi$-fields in $H(\phi)$. This completes the demonstration 
of our theorem.  Note that the ``frozen''
variables ${\bar \psi}_{\bf i}$ act like external fields 
in ${\bar H}(\eta)$ which do not break the global
symmetry group of transformations ${\bf U}_{jk}$.

{\it Corollary I: Elitzur's theorem.} Any local quantity 
(i.e. involving only a finite number of fields)
which is not gauge invariant under a local or zero-dimensional 
gauge group has a vanishing mean value at 
any finite temperature. This is a direct consequence of 
Eq.(\ref{mast}) and the fact that ${\bar H}(\eta)$ 
is a zero-dimensional Hamiltonian.

{\it Corollary II.} Any local quantity which is not gauge 
invariant under a one-dimensional gauge group
has a vanishing mean value at any finite temperature for 
systems with finite range interactions. This 
is a consequence of Eq.(\ref{final}) and the absence of phase 
transitions for one-dimensional Hamiltonians
with finite range interactions. Note that 
${\bar H}(\eta) \equiv H({\bar \psi},\eta)$ is a one of these Hamiltonians
and $f(\eta_{\bf i})$ is a non-invariant quantity under 
the global symmetry group $G_j$ [see Eq.(\ref{nonin})],
\cite{discrete}.

{\it Corollary III.} Any local quantity which is not gauge invariant 
under a two-dimensional continuous gauge 
group has a vanishing mean value at any finite temperature for 
systems with finite range interactions. This results 
from the combination of our theorem [Eq.(\ref{final})] 
with the Mermin-Wagner theorem:
\begin{equation}
lim_{h\rightarrow 0} lim_{N \rightarrow \infty} 
\frac {\sum_{\{ \eta_{\bf i} \} } f(\eta_{\bf i})  
e^{-\beta H({\bar \psi},\eta)-\beta h \sum_{{\bf i} 
\in {\cal C}_j}  \eta_{\bf i}}}{z_{{\{ {\bar \psi} \}}}} = 0.
\end{equation}
Where we have used that $G_j$ is a continuous symmetry 
group of ${\bar H}(\eta)= H({\bar \psi},\eta)$, 
$f(\eta_{\bf i})$ is a non-invariant quantity for 
$G_j$ [see Eq.(\ref{nonin})], and ${\bar H}(\eta)$ is a 
two-dimensional Hamiltonian that only contains finite range 
interactions.
\cite{Bog}  

The generalization of our theorem to the quantum case is very 
straightforward if we choose a basis of eigenvectors 
of the local operators that are linearly coupled to the symmetry 
breaking field $h$. In this basis, the states can 
be written as a direct product 
$ |\phi \rangle = | \psi \rangle \otimes | \eta \rangle$. 
Eq.(\ref{final}) is re-obtained 
with the sums replaced by traces over the states $| \eta \rangle$:
\begin{equation}
|\langle f({\boldsymbol \phi}_{\bf i}) \rangle_{h,N}| \leq \mid   
\frac {{\rm Tr}_{\{ \eta_{\bf i} \} } f({\boldsymbol \eta}_{\bf i})  
e^{-\beta H({\bar \psi},{\boldsymbol \eta})-\beta h \sum_{{\bf i} 
\in {\cal C}_j}  {\boldsymbol \eta}_{\bf i}}}
{{\rm Tr}_{\{ \eta_{\bf i} \} }  
e^{-\beta H({\bar \psi},{\boldsymbol \eta})-\beta h \sum_{{\bf i} 
\in {\cal C}_j}  {\boldsymbol \eta}_{\bf i}}} \mid, 
\label{quantum}
\end{equation}
In this case, $| {\bar \psi} \rangle$ corresponds to one particular 
state of the basis $| \psi \rangle$ that maximizes
the right side of Eq.(\ref{quantum}). Generalizing standard proofs,
e.g. \cite{assa}, we find in the quantum arena
a zero temperature extension of 
Corollary III whenever there is a gap in the excitation spectrum:

{\it Corollary IV.} If a gap exists
in a system possessing a $d \le 2$ dimensional continuous symmetry
(or, at least, an {\em emergent} continuous symmetry) 
in its low energy sector, then the expectation value of 
any local quantity which is not invariant under this symmetry, 
strictly vanishes at zero temperature.

\section{Physical Realizations and Implications} 

Armed with the above results, we briefly review 
known consequences of Elitzur's theorem and 
then focus at length on several of the applications 
of our new results.

\subsection{Local gauge symmetries- Pure and Matter coupled gauge
theories and Spin glass systems}

a) {\it Gauge theories}-
The application of Elitzur's theorem in 
gauge theories is well known and reviewed
in excellent works such as \cite{Kogut}.
No work on Elitzur's theorem is complete 
without a mention of its most prominent application.
In theories of matter at lattice sites ($\sigma_{i}$) coupled 
to gauge fields ($U_{ij}$) residing on links, the
action is a sum of a plaquette product of the gauge
fields and (minimal) coupling between matter fields through
the gauge \cite{Wegner}. To illustrate, consider the action in 
a $Z_{2}$ setting,
\begin{eqnarray}
S = - K \sum_{\Box} U_{ij}U_{jk}U_{kl}U_{kl}
- J \sum_{ij} \sigma_{i} U_{ij} \sigma_{j},
\end{eqnarray}
with $U_{ij} = \pm 1$ and $\sigma_{i} = \pm 1$ elements of $Z_{2}$.
Many pioneering results were found 
by \cite{Fradkin}. The action is invariant under
the local gauge transformation
$\sigma_{i} \to \eta_{i} \sigma_{i}, U_{ij} \to \eta_{i} U_{ij} 
\eta_{j}$ with $\eta_{i} = \pm 1$ an element of
$Z_{2}$. Defining the gauge invariant link variables
$z_{ij} \equiv \sigma_{i} U_{ij} \sigma_{j}$ of which the 
action is a functional, we note that
any correlator involving a product of any number of $z$'s
(``mesons'') need not vanish. In the absence of coupling to matter
($J =0$)- the ``pure gauge theory''-  
only products of $U_{ij}$ along closed ``Wilson'' loops 
($W = \langle \prod_{ij \in C} U_{ij} \rangle$) are
gauge invariant and may attain finite 
expectation values. 

b) {\it Spin glasses}- 
Perhaps one of the simplest realizations of spin-glasses is the 
Edwards-Anderson (EA) model \cite{EA} with 
\begin{eqnarray}
H = - \sum_{\langle ij \rangle} J_{ij} \vec{S}_{i} \cdot \vec{S}_{j}  - \vec{h} \cdot 
\sum_{i} \vec{S}_{i}.
\label{EA.}
\end{eqnarray}
Here, $i$ and $j$ are nearest neighbor sites of a regular lattice and
the three component spins $\vec{S} = (S_{x},S_{y},S_{z})$ 
are considered as classical vectors. In
the EA model, the exchange constants $J_{ij}$ are independent random 
variables drawn from a (Gaussian) probability distribution $P(J_{ij})$.
In spin glass systems, the experimentally measured expectation 
value of any observable corresponds to a quenched 
average over the distribution $P(J)$, 
\begin{eqnarray}
\langle f(\{\vec{S}\}) \rangle_{quench} = \int dJ P(J) \langle   
f(\{\vec{S}\}) \rangle_{J}.
\label{quench}
\end{eqnarray}
Here, $\langle f(\{\vec{S}\}) \rangle_{J}$ denotes the expectation
value of $f(\{\vec{S}\})$ for a given {\em quenched}
distribution of exchange constants $\{J_{ij}\}$.
The basic premise of spin glass systems is
that the free energies (and their derivatives)
are to be averaged via the distribution $P(J)$
to obtain the corresponding quenched quantities.
The quenched expectation value 
of Eq.(\ref{quench}) is, for any even 
distribution $P(J)$, 
invariant under the following local $Z_{2}$ gauge transformation,
\begin{eqnarray}
\vec{S}_{i} \to \eta_{i} \vec{S}_{i}, ~J_{ij} \to \eta_{i} J_{ij} \eta_{j},
\label{Z2SG}
\end{eqnarray}
with $\eta_{i} = \pm 1$ (an element of $Z_{2}$). \cite{mattis}
Performing the quenched disorder 
average of Eq.(\ref{quench}) over $J_{ij}$ is a central task 
in spin glass problems and ingenious schemes
have been devised. Parisi showed
that the quenched average may be done by introducing $n$ (with $n \to 0$)
replicas of the spin field $\{\vec{S}_{i}^{a}\}_{a=1}^{n}$ at each
site all coupling via the same $J_{ij}$ thus still satisfying 
the gauge transformation of Eq.(\ref{Z2SG})
detailed above with a replica independent
$\eta_{i}$ \cite{explain_replica}.  Within the replica formulation, 
the average over $J_{ij}$ may be performed first 
leading to a functional of the replicated spin
fields alone. To attain a finite expectation
value, the quantity of interest must be gauge invariant.
The lowest order quantity invariant under local transformation
is \cite{Parisi}, \cite{parisi94} 
$Q^{ab}_{i} = \langle \vec{S}_{i}^{a} \cdot \vec{S}_{i}^{b} \rangle$
(as both spin field suffer a factor of $\eta_{i}$
under the local gauge transformation, and $\eta_{i}^{2} =1$).
In this manner, we see how the usual spin-glass
overlap order parameters ($Q^{ab}_{i}$
and their sums) are dictated by Elitzur's  
theorem for the n-replica action. 
c) Many other examples are found in numerous
spin systems of current interest, For instance, Kitaev's \cite{kitaev}
exact solution of a special spin system on a honeycomb
lattice owes its existence to gauge-like local symmetries
implemented via spin products around individual hexagons.
When these local symmetries are fused with 
Elitzur's theorem, we find
that on-site magnetization is precluded.

\subsection{Exact Higher-Dimensional Gauge-like
symmetries: Orbital systems}  

a) {\it Spins}- In 3d orbital systems
on cubic lattices, each transition metal
atom is surrounded by 
an octahedral cage of oxygens. Crystal field 
splitting lifts the degeneracy of the five 3d orbitals
of the transition metal 
to that of two higher energy $e_{g}$ 
levels (residing in the space spanned by the two
orbital states
$|d_{3z^{2}-r^{2}} \rangle$ and $| d_{x^{2}-y^{2}} \rangle$
wherein the electronic wave-functions point towards
the surrounding oxygens with an associated high energy
Coulomb penalty)
and to three lower energy $t_{2g}$ levels
(spanned by the three states $| d_{xy} \rangle$, $| d_{xz} \rangle$
and $| d_{yz} \rangle$ in which the wave-functions
point away from the surrounding 
oxygen atoms). Following \cite{Harris},
the latter three $t_{2g}$ states will
henceforth be denoted by $|X \rangle, | Y \rangle,$ 
and $|Z \rangle$. 

\begin{figure}[htb]
\vspace*{-0.5cm}
%\vspace*{0cm}
\includegraphics[angle=-90,width=8cm]{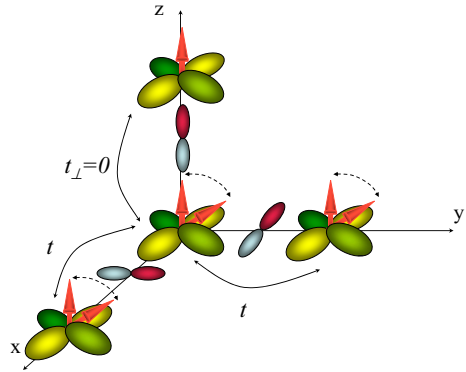}
\vspace*{-0.0cm}
\caption{ (color online)
The anisotropic hopping amplitudes leading to 
the Kugel-Khomskii (KK) Hamiltonian (Eq.(\ref{KKe})). Similar
to \cite{Harris}, 
the four lobed states denote the $3d$ orbitals
of a transition metal while the intermediate small $p$ orbitals
are oxygen orbital through which the super-exchange process 
occurs. The dark and bright shades 
denote positive and negative regions
of the orbital wave-function. Due to orthogonality with the intermediate 
oxygen $p$ states, in 
any orbital state $|\alpha \rangle$ (e.g. $| Z \rangle 
\equiv | d_{xy} \rangle $ above), hopping is 
forbidden between two states separated along the cubic 
$\alpha$ (Z above) axis. The ensuing super-exchange (KK) Hamiltonian
exhibits a two-dimensional $SU(2)$ gauge symmetry
corresponding to a uniform rotation of all
spins whose orbital state is $|\alpha \rangle$ 
in any plane orthogonal to the cubic direction $\alpha$. In the above
all spins whose orbital states are $| Z \rangle $ have been rotated
in an arbitrary plane orthogonal to the cubic Z axis.}
\label{figu}
\end{figure}

The super-exchange 
Kugel-Khomskii Hamiltonian 
(KK) \cite{KK} depicts a coupling between orbital and spin degrees
of freedom in transition metal 3d systems such as the $t_{2g}$ 
titanate LaTiO$_{3}$ and the vanadate LaVO$_{3}$. 
As demonstrated by KK, 
in orbital systems the super-exchange 
Hamiltonian is anisotropic with a strong dependence on the orbital state.
The overlap between the transition 
metal orbitals amongst 
themselves (and more notably via intermediate
oxygen orbitals) strongly 
determines the strength (and
viability if any) of the spin exchange 
interactions along the crystalline
directions. In $e_{g}$ systems, 
the orbital component of the KK Hamiltonian exhibits, on a cubic 
lattice of size $L \times L \times L$, a discrete
$[Z_{2}]^{3L}$ ($d=2$) gauge-like symmetry corresponding
to reflections of spins in entire planes about
certain axis \cite{NBCv}, \cite{BCN}. 
As emphasized in  \cite{Harris},
the three level $t_{2g}$ version of 
the KK Hamiltonian, which is relevant to the titanates, 
prohibits direct hopping
of electrons of a certain orbital flavor along
one lattice direction. The key feature of
the super-exchange Hamiltonian is
that it prohibits hopping via intermediate 
oxygen p orbitals between any two electronic
states of orbital flavor $\alpha$ ($\alpha = X, Y$, or $Z$)
along the $\alpha$ axis of the cubic lattice (see Fig.\ref{figu}).

In these materials the 
KK Hamiltonian \cite{KK} can be written as
$H = H_{X} + H_{Y} + H_{Z}$ with 
\begin{eqnarray}
H_{\alpha} =  J \sum_{\langle ij \rangle \in \alpha}
~ \sum_{\beta, \gamma \neq \alpha} ~ \sum_{\sigma \eta}
c^{\dagger}_{i, \beta, \sigma} c_{i,\gamma, \eta} 
c^{\dagger}_{j, \gamma, \eta}
c_{j,\beta,\sigma}. 
\label{KKe}
\end{eqnarray}
In this Hamiltonian, $\langle i j \rangle \in \alpha$
is a nearest neighbor bond along the cubic lattice $\alpha$
axis, $c^{\dagger}_{i \beta \sigma}$ creates an electron
at the cubic lattice site $i$ in a $\beta$ orbital state
of spin $\sigma$.

As is evident from Eq.(\ref{KKe}), a 
uniform rotation of all spins, whose electronic orbital state
is $|\alpha \rangle$, in 
any given plane ($P$) orthogonal  
to the $\alpha$ axis \cite{Harris}
\begin{eqnarray}
c^{\dagger}_{i \alpha \sigma} = \sum_{\eta} U^{(P)}_{\sigma, \eta}
d^{\dagger}_{i \alpha \eta}
\end{eqnarray}
with $\sigma, \eta$ the internal spin polarization
directions, leaves the KK Hamiltonian invariant.

This leads to
a conservation of the net spin of the electrons
of orbital flavor $|\alpha \rangle$ 
in any plane orthogonal to the cubic $\alpha$ axis. 
This manifests itself 
via a continuous $SU(2)$ planar gauge group.
The continuous planar 
rotation 
$\hat{O}_{P;\alpha} \equiv [\exp(i\vec{S}^{\alpha}_{P} \cdot 
\vec{\theta}^{\alpha}_{P})/\hbar]$ 
is generated by $\vec{S}^{\alpha}_{P} = \sum_{i \in P} \vec{S}_{i}^{\alpha}$,
the sum of all the spins $\vec{S}^{i, \alpha}$ in the orbital state
$\alpha$ in any plane $P$ 
orthogonal to the direction $\alpha$, is 
a symmetry operation (see Fig.\ref{figu}). Our theorem immediately 
prohibits a finite
magnetization in these systems $\langle \vec{S} \rangle =0$
(Corollary III)- coinciding with the 
central result of \cite{Harris}. 
We note that this symmetry does not prohibit the much more robust 
nematic spin ordering. For instance, in the presence of
orbital ordering in the $| \alpha \rangle$ state,
the order parameter
$P \equiv \langle \vec{S}_{\vec{r}}  \cdot
\vec{S}_{\vec{r}+ \hat{e}_{\eta}} \rangle$
along the $\eta =$ x,y, or z axes
with $\eta \neq \alpha$
need not vanish by our theorem.
In real systems, the ideal mathematical model of KK
with its perfect gauge-like symmetry and Elitzur theorem
forbidden magnetization is likely corrupted by small effects
(e.g. the hopping amplitude $t_{\perp}$
along the $\alpha$ axis does not identically vanish). 
Indeed finite temperature magnetization is seen in many
of these compounds.
Generically, in systems where the gauge-like symmetries inhibiting order 
are lightly lifted, we may naturally anticipate the 
now allowed ordering to be fragile and to
exhibit a low transition temperature.

b) {\it Orbitals}- Orbitals in 3d systems 
interact amongst themselves via Jahn-Teller distortions
which for much the same reasons as in the KK Hamiltonian are anisotropic. 
The canonical prototype of all orbital interactions
is the orbital compass model \cite{Brink03}. 
To streamline the essential physics
in its simplest form, we examine the two-dimensional version
of the orbital compass model. A complete mathematical analysis 
is detailed elsewhere \cite{NBCF}.
In the two-dimensional orbital compass model
at each site~$\vec{r}$ of an $L \times L$  
square lattice there is a S=1/2 operator 
denoted, throughout for XY spins, 
by a lower case 
~$\vec{s}_{\vec{r}} = \frac{\hbar}{2} \vec{\sigma}_{\vec{r}}$
(the upper case $\vec{S}$ will be reserved
for three component spins). 
The planar orbital compass model Hamiltonian is:
\begin{equation}
\label{Hcompass}
H=\frac{J}{4} 
\sum_{\vec{r}} (\sigma^{x}_{\vec{r}} \sigma^{x}_{\vec{r}+\hat{e}_{x}} 
+ \sigma^{y}_{\vec{r}} \sigma^{y}_{\vec{r}+\hat{e}_{y}}).
\end{equation}

In the three-dimensional model, additional nearest neighbor
$\sigma^{z} \sigma^{z}$ bonds appear parallel to the z-axis. We will
come to this in Eq.(\ref{3Dcompass}) later on.
To tabulate the symmetries in this system,
we define an operator on an arbitrary line P (of intercept y)
$\hat{O}_{P;y} \equiv \prod_{\vec{r} \in P} \sigma^{y}_{\vec{r}}$ 
with a similar definition for $\hat{O}_{P;x}$. Up to global
multiplicative phase factors, $ \hat{O}_{P;y}$
is a rotation by $\pi$ about the x axis 
of all sites $\vec{r}$ in the line $P$ 
whose x component is $r_{x} = x $. Similarly,
the operator $\hat{O}_{P;x} =  \exp[i (\pi/2) \sigma^{x}_{P}]$
with $\sigma^{x}_{P} = \sum_{\vec{r} \in P} \sigma^{x}_{\vec{r}}$.
This is very much similar to the operators which appeared
in our discussion above of the planar gauge-like symmetries of
the KK model, yet for the orbitals here, these symmetries
are discrete (the angle of rotation $\theta_{P}$ is quantized).
From the products $\hat{O}_{P;z}^{-1} \sigma^{P;x,y}_{\vec{r}} 
\hat{O}_{P;z} = - \sigma^{x,y}_{\vec{r}}$, it is clear that  $\hat{O}_{P;z}$
%$\hat{O}_{P;z}^{-1} \sigma^{z}_{\vec{r}} \hat{O}_{P;z} =  
%\sigma^{z}_{\vec{r}}$ 
is a symmetry of $H$ .
Similarly, $\hat{O}_{P;x}$ inverts the y-component of all spins on the line
of intercept $x$ while leaving $\sigma^{x}$ untouched. 
In arbitrary spin size (S) versions of this model, the
string operators above containing spin-1/2 Pauli matrices
may be replaced by the corresponding spin-S $\pi$-rotations 
of all spins in the line $P$ .
These ``string'' operators spanning spin operators on entire lines
commute with the Hamiltonian, $[H, \hat{O}_{P;\alpha}]=0$.
As in the case of spins in orbital $t_{2g}$ systems,
rotations of individual lower-dimensional ``planes'' (lines) 
about an axis orthogonal to them
leave the system invariant.
As a consequence of these symmetries, 
each state is, at least, ${\cal{O}}(2^{2L})$ degenerate. 

As there exist one-dimensional gauge-like symmetries
$\hat{O}_{P;x,y}$ in this system, by corollary (ii)
the magnetization vanishes identically and only quantities
invariant under $\hat{O}_{P;x,y}$ 
may attain a finite expectation value. Indeed, it has 
been established that a nematic-type order 
respecting the gauge-like 
symmetries exists in related classical 
extensions of this system \cite{Mishra,NBCv,BCN}.
More sophisticated 
order parameters $\langle T \rangle$
(invariant under $\{ {\hat{O}}_{P;x,y} \}$)
and their composites may be trivially constructed \cite{NBCF}. 
In these, $\langle T \rangle \equiv \langle 
\prod \sigma^{x}_{\vec{r}} \sigma^{y}_{\vec{r}^{\prime}} \rangle$,
the number of $\{ \sigma^{x}_{\vec{r}} \}$
on any given horizontal row and the number 
of $\{ \sigma^{y}_{\vec{r}^{\prime}} \}$ on any
given vertical column are both even.
Due to the anti-commuting nature of $\sigma^{x}_{\vec{r}}$
and $\sigma^{z}_{\vec{r}}$, any specific operator $T_{i}$ commutes with all
$2L$ inversion operators $\{ {\hat{O}}_{P;x,y} \}$,
with the Hamiltonian $H$ and 
with any other operator $T_{j \neq i}$ \cite{NBCF}.
Obviously, the largest and most physically
relevant order parameter is the minimal
gauge invariant dimer like nematic order
parameters $\langle S^{\alpha}_{i} S^{\alpha}_{i + \hat{e}_{\alpha}} 
\rangle$ (with $\alpha = x,y$), and 
their linear combinations.

For the three-dimensional ``$120^{\circ}$ model'' 
encapsulating the Jahn-Teller distortions and the
orbital only component of the orbital-spin KK 
Hamiltonian in 
``$e_{g}$'' orbital systems (such as e.g. the colossal magneto-resistive
manganate LaMnO$_{3}$,  the cuprate KCuF$_{3}$,
the nickelate LiNiO$_{2}$, Rb$_{2}$CrCl$_{4}$,
and many other systems)  
novel discrete planar gauge-like symmetries 
and a finite on-site magnetization were found 
\cite{NBCv},\cite{BCN}. The three-dimensional 
``$120^{\circ}$ model'' model possesses an exact discrete ($d=2$) 
$[Z_{2}]^{3L}$ gauge-like symmetry 
(by sequential planar Rubick's cube like reflections
about internal spin directions;
See Fig.1 in Refs. \cite{NBCv},\cite{BCN} for an illustration). 
Unlike the case for the spins in the KK Hamiltonian for orbital 
$t_{2g}$ systems, 
there are no exact nor emergent continuous 
gauge-like symmetries. In accord with
our postulates, the three-dimensional $120^{\circ}$ model
can (and indeed does) have a finite magnetization \cite{NBCv}, \cite{BCN}. 

\subsection{Exact Gauge-Like Symmetries: Superconducting Arrays} 

A related system having only four-spin interactions
(i.e. with absolutely no two spin interactions), 
which emulates $(p+ip)$ superconducting grains 
(such as those of e.g. Sr$_{2}$RuO$_{4}$) 
on a square grid, was recently investigated by Xu and Moore 
 \cite{Xu03}, \cite{Xu04}.
The Hamiltonian of this system is 
\begin{eqnarray}
H = - K \sum_{\Box} \sigma^{z} \sigma^{z} \sigma^{z} \sigma^{z}
- h \sum_{\bf r} \sigma_{\bf r}^{x}.
\label{XM}
\end{eqnarray}
Here, the four spin product is the product of
all spins common to a given plaquette $\Box$. Note that
the spins reside on the vertices on the plaquette
(not on its bonds as gauge fields).
It has been established via duality mappings \cite{NBCF}, 
that this model of superconducting arrays is identical to 
the two-dimensional orbital compass model discussed in III{\bf B}b).
Not too surprisingly, the systems investigated by Xu and Moore 
have clear one-dimensional gauge-like symmetries similar to those
of the two-dimensional orbital compass discussed in III{\bf B}b).
Rather explicitly, 
\begin{eqnarray}
\hat{O}_{P} = \prod_{\vec{r} \in  P} \sigma^{x}_{\vec{r}}
\end{eqnarray}
with $P$ any horizontal row or vertical column,
is a symmetry of Eq.(\ref{XM}).
These symmetries are trivially related to
the gauge-like symmetries of the planar 
orbital compass model upon
invoking the duality mappings of 
\cite{NBCF}.

\subsection{Other systems with Gauge-Like Symmetries}

Many other pioneering works unveiled plenty of
new gauge-like symmetries in other systems. In \cite{Arun02},
it was shown that ring exchange Bose metals,
in the absence of nearest neighbor Boson
hopping, exhibit much the same one-dimensional
gauge-like symmetries discussed above.
Similar {\em sliding symmetries} of 
Hamiltonians (actions) invariant under 
arbitrary {\em continuous} deformations of a certain
field along a transverse direction,
\begin{eqnarray}
\phi(x,y) \to 
\phi(x,y) + f(y),
\label{slide}
\end{eqnarray} 
appear in many systems. As much throughout this work, in these systems 
there exists no rigidity to shear deformations
along one (or more) direction(s) ($y$ in Eq.(\ref{slide})).
Many such systems were discovered
and discussed at length in rather 
novel works on Quantum Hall liquid crystalline phases
\cite{lawler}, \cite{rad}, a number of models of lipid 
bilayers with intercalated DNA strands \cite{ohern},
and ``Luttinger Liquid Arrays'' or "sliding Luttinger liquids" 
\cite{emery2000} which, as  
suggested in various works, may be intimately related to 
``stripes'' \cite{stripea},\cite{stripeb}. 
The consequences of our theorem (and Corollaries II, III in particular)  
are immediate for these systems, $\langle \phi \rangle = 0$.

\subsection{Exact Gauge-Like Symmetries by Inverted
Dimensional Reductions} 

Many spin systems  
can be systematically generated to have the gauge-like symmetries of
orbital systems (as well as many others) \cite{Zohar}. 
We found a very simple algorithm
doing so which very lucidly invokes dimensional reduction in the 
reverse order. The basic idea underlying our 
scheme is that a high-dimensional system containing
decoupled chains or planes is, of course, 
one-dimensional and possesses trivial one- or two-dimensional
gauge-like symmetries. To illustrate, we first consider a one-dimensional 
XY spin chain having nearest neighbor interactions,
\begin{eqnarray}
H = J \sum_{\langle i j \rangle} \vec{s}_{i} \cdot \vec{s}_{j}, 
\end{eqnarray}
with $i$ and $j$ nearest neighbor lattice sites
and $\vec{s} = (s_{x}, s_{y})$ an XY spin. 
This translationally invariant Hamiltonian may, of course, be expressed in Fourier
space, 
\begin{eqnarray}
H = \frac{1}{2N} \sum_{\vec{k}} s_{\alpha}(\vec{k}) v_{\alpha \beta}(\vec{k}) 
\label{reduce}
s_{\beta}(-\vec{k}),
\end{eqnarray}
where $\alpha$ and $\beta$ denote the two internal
spin polarization directions, and the matrix $v_{\alpha \beta} =
(2J \delta_{\alpha \beta} \cos k)$. Let us now
make this one-dimensional system
higher-dimensional by replacing the scalar quantity
$k$ by more complicated quantities.
If we replace the scalar momentum $k$ of the 
one-dimensional spin chain by a two-dimensional $\vec{k}$
which appears in $v(\vec{k})$ only via 
$[\cos (\vec{k} \cdot \vec{a})]$ with $\vec{a}$ a 
real vector, then the two-dimensional system will be 
a trivial realization of decoupled chains (interacting along the 
direction $\hat{a}$) and trivially possessing one-dimensional
gauge symmetries. We may now replace the c-number coefficients
$a_{i}$ by operators to obtain non-trivial systems. 
Some of the simplest examples are afforded when $\{a_{i}\}$
are chosen to be projection operators.
For instance, by replacing the scalar $k$ 
of the one-dimensional spin chain by a simple
two-dimensional diagonal matrix, we obtain 
$v(\vec{k}) = 2J \diag(\cos k_{x}, \cos k_{y})$.
(Here and throughout, $\diag(a_{1}, a_{2}, ..., a_{d})$
refers to a $d$-dimensional diagonal whose diagonal elements
are $\{a_{1}, a_{2}, ..., a_{d} \}$). 
In the $v(\vec{k})$ 
above, we replicate the single momentum coordinate to generate
a higher-dimensional interaction.
Insertion of this matrix into Eq.(\ref{reduce}),
leads to the two-dimensional orbital compass model
of {\bf B}b) having $[Z_{2}]^{2L}$ gauge-like symmetries
of ``en-block'' rotations by $\pi$ of all spins 
belonging to a certain row/column.

Similarly,
by setting $v(\vec{k}) = 2J \diag(\cos k_{x}, \cos k_{y}, \cos k_{z})$
in Eq.(\ref{reduce}) for a three component spin system 
$\vec{S}= (S_{x},S_{y},S_{xz})$,
we obtain the three-dimensional orbital compass model (on a cubic
lattice of size $L \times L \times L$)
\begin{eqnarray}
H = J \sum_{i} (S_{i}^{x} S_{i+\hat{e}_{x}}^{x}  + S_{i}^{y} 
S_{i+\hat{e}_{y}}^{y} + S_{i}^{z} 
S_{i+\hat{e}_{z}}^{z}).
\label{3Dcompass}
\end{eqnarray}
This system has an exact $[Z_{2}]^{3L^{2}}$ symmetry for classical spins
(along each chain parallel to the cubic $\alpha$ axis,
we may reflect $S_{\alpha} \to - S_{\alpha}$
yet keep all other spin polarizations unchanged,
$S_{\beta \neq \alpha} \to S_{\beta \neq \alpha}$)
as well as a lower exact $[Z_{2}]^{3L}$ gauge-like 
symmetry (forming a subset of the
larger $[Z_{2}]^{3L^{2}}$ symmetry
present for classical spins) when the spins are quantum:
in this case we may rotate all spins in a plane orthogonal to
the cubic lattice direction $\alpha$ by $\pi$ about the 
internal $S_{\alpha}$ quantization axis.

In Eq.(\ref{reduce}) for an XY spin system, we
may further  generate interactions
along arbitrary tilted rays by considering
$v= 2J \cos(k_{x}+ \lambda k_{y} \sigma_{3})$
with $\sigma_{3}$ the Pauli matrix. An insertion
this kernel into Eq.(\ref{reduce}) leads
to real space interactions of the 
x components of the spins along chains
parallel to the $(1, \lambda)$ direction
and of an exchange interaction amongst the 
y components of the spins along the 
$(1, -\lambda)$ direction. 
By tuning the $\lambda \to 0$, the
two ``clapping'' $(1, \pm \lambda)$ 
directions fold back and degenerate into
the original one-dimensional XY spin chain. 
We find that for all $\lambda \neq 0$, the classical symmetries
are much the same as for the two-dimensional orbital
compass model: reflecting all $s_{x} \to - s_{x}$
while leaving $s_{y}$ untouched along any chain 
parallel to $(1, \lambda)$. Thus, according to our theorem
and corollary (II), no finite on-site magnetization
is possible in classical XY systems with the
above Hamiltonian. Nematic like order is possible
and was indeed found \cite{Mishra} for $\lambda = \pm 1$. 
For $\lambda \neq  \pm1 $, there is, however,
no canonical transformation that may be applied
to the quantum spin model to encapsulate these 
symmetries. 

If we start with a nearest neighbor 
three component (Heisenberg) spin model 
on the square lattice and replace the usual isotropic interaction 
kernel $2J \diag (\cos k_{x} + \cos k_{y},\cos k_{x} + \cos k_{y})$
by $v= \diag(\cos k_{x} + \cos k_{y}, 
\cos k_{x} + \cos k_{y}, \cos k_{z})$ to generate
a spin system residing on a cubic lattice, we obtain
a new system with many exact gauge-like symmetries. In classical systems, 
for all chains parallel to the z-axis we may take $S_{z} \to -S_{z}$
and for all planes parallel to the XY plane we may perform a 
uniform continuous rotation about the z axis- 
a $[U(1)]^{L} \times Z_{2}^{L^{2}}$ symmetry. 
If the spins are quantum, only the canonical $[U(1)]^{L}$
symmetries are present. Our theorem precludes 
finite on-site order for both classical and
quantum spins on a cubic lattice with the above 
spin Hamiltonian.

We may similarly start with many other lower-dimensional
systems to generate higher-dimensional systems by these
operations. In some instances, 
these transformations lead to systems of intense physical interest
(e.g. orbital compass models employed for elucidating the
properties 3d transition metal compounds at large \cite{Brink03}, 
and crystal field split $t_{2g}$ systems 
in particular). For each of these systems, our theorem and its corollaries
can be applied to often inhibit the possibility
of a finite on-site magnetization. It should
be emphasized that although these
models are generated by artificially ``lifting''
a low-dimensional momentum,
the resulting Hamiltonians are, obviously, not lower dimensional. 
We are only considering symmetry properties
here and not the thermodynamics.
In the current context, ``dimensional reduction''
may be understood from a technical
point of view as an algorithm for 
yielding gauge-like symmetries
in a natural fashion. Thermodynamic
dimensional reduction of
these models occurs only
in the large $n$ (spherical)
limit of classical renditions
of all of the above mentioned spin systems
(e.g. the two-dimensional compass model
is identical to an XY chain).
We will briefly comment to other special properties 
of large $n$ systems when 
discussing new emergent gauge-like 
symmetries that we found in Brazovskii-type systems
relevant, amongst other things, to liquid crystal systems, 
theories of structural glasses, electronic systems
favoring phase separation that is inhibited by
long range Coulomb interactions, and to many other arenas.

By reverse engineering various given gauge-like lattice
symmetries to find corresponding Hamiltonians, 
we found additional models on different lattices having 
gauge-like symmetries when polarization (spin space) 
dependent spin-spin interactions were slaved to 
the bond directional on the lattice (external space)
to produce many different real space interaction kernels 
$v_{\alpha \beta}(\vec{k})$ (or their direct
real space transforms $V_{\alpha \beta}(\vec{r}_{i}, \vec{r}_{j})$).

\subsection{Emergent gauge-like symmetries in spin, orbital, and 
other systems}

We now examine situations wherein a gauge-like symmetry {\em emerges}
within the low energy sector. In some instances, it is  
present in leading order fluctuations about the classical ground state
(e.g. $1/S$ spin wave corrections). From corollary (IV), 
we know that in the presence of a gap, expectation 
values of any symmetry non-invariant quantities must
vanish. In general systems having 
various emergent gauge-like symmetries, non-gauge
invariant quantities may often vanish.  I. Quantum systems:
a) Geometrically Frustrated Spin systems: In many spin systems 
(e.g. the kagome, pyrochlore, 
and checkerboard magnets), the availability of
many spins vis a vis the number of building
blocks on which they reside allows the proliferation
of many low energy states \cite{moessner}. Pyrochlore magnets 
such as the S= 3/2 spinel ZnCr$_{2}$O$_{4}$ have remarkable magnetic
behavior (such as e.g. very clearly detectable
spin waves linking degenerate states). The checkerboard 
lattice \cite{oleg} may be viewed as a two-dimensional projection 
of the three dimensional pyrochlore lattice 
composed of corner sharing tetrahedra on the plane.
The system is composed of nearest neighbor bonds 
on the square lattice augmented by additional
diagonal, next nearest neighbor, interactions
along all diagonals of one plaquette sub-lattice
(hence the name ``checkerboard''). Both the 
pyrochlore and checkerboard magnets 
has an exponential in volume classical 
ground state degeneracy as in other 
systems of emergent local (or $d=0$) gauge symmetry.  
Leading order $(1/S)$ corrections were computed finding the 
effective spin-wave Hamiltonian
for fluctuations about the ground states \cite{oleg,henley}. 
Much like the orbital systems discussed
in Section (IIIBb), the resulting Hamiltonian
possesses one dimensional gauge like
symmetries.  In \cite{assa03}, an effective low 
energy Hamiltonian was derived
for these quantum systems without the
need for spin-wave theory.  The resulting
effective Hamiltonian  
looks much the same as the KK Hamiltonian
for orbital systems 
discussed earlier and displays much the same
discrete gauge-like symmetries
seen in the KK relevant to ordering
in $e_{g}$ materials \cite{NBCv},\cite{BCN}.
Gauge-like theories 
for geometrically frustrated magnets
with string operators generating 
freely propagating ``spinons''
are also found elsewhere, e.g.
\cite{fisher} for the pyrochlore
magnet. The large $S$ ground states of
kagome antiferromagnets may be linked to
each other via all possible contortions
of a two-dimensional membrane. 
The effective low energy Hamiltonian of the quantum 
Kagome magnet is, once again, similar to 
the KK Hamiltonian. \cite{assa04}

b) Spin systems with 
ring exchange: In a recent work
\cite{BT}, emergent one dimensional gauge like
symmetries were found in a quasi-exactly solvable square 
lattice spin system of a Heisenberg antiferromagnet augmented 
by ring exchange interactions
of comparable strength. Here, the
presence of one-dimensional gauge like
symmetries went hand in hand with fractionalization:
spinons propagating along the diagonal directions
incurred no energy penalty while
spin motion in other directions led to
a confining potential. In an upcoming work \cite{prep}, 
we identify gauge-invariant order parameters
in this system and further suggest the
existence of a finite temperature ``order
out disorder'' transition suggesting
natural finite temperature fractionalization
in these two-dimensional systems.

II. Classical spin systems:
a) Within the  
classical (large S) limit of d=3 orbital compass of Eq.(\ref{3Dcompass})
(relevant to such compounds as YVO$_{3}$
LaVO$_{3}$, and LaTiO$_{3}$), we find an
emergent continuous rotational symmetry of all
the spins in an entire plane about a spin
axis orthogonal to the plane direction.
For instance, taking any uniform spin state (automatically a ground 
state of Eq.(\ref{3Dcompass})) and rotating all 
of the spins in a given plane orthogonal to the z-axis
(i.e. all spins on sites $\vec{r}$ satisfying $r_{z} = z$)
 by an arbitrary angle $\theta$ about the 
$S_{z}$ spin polarization axis leads to another ground state.
This can be further mutated by similar operations
in other planes (parallel and orthogonal).
Compounding all of the symmetries
together, this corresponds to an emergent 
$[U(1)]^{3L}$ degeneracy.
Indeed, it was found (\cite{BCN}) that
the system has nematic like
order invariant under this continuous
planar gauge-like symmetry. b) The Fully Frustrated XY model on
a honeycomb lattice: within the ground state of such an XY system
having half a fluxon threading each hexagonal ``plaquette''
\cite{korshunov}, each ``plaquette'' has exactly two
bonds having zero phase gradient (similar
to the requirement of one singlet per
square plaquette in \cite{BT})
with all other bonds having a 
phase gradients of $\pi/4$.
Connecting the zero gradient
bonds to each other leads to non-intersecting stripes
threading the system. There are several ways of 
generating such stripe configurations,  
all of which are inter-related by an emergent
one dimensional like gauge symmetry.
c) An Ising spin systems on a square lattice with competing 
next and next nearest neighbor \cite{zn01} 
interactions, having a shell of minimizing modes
in Fourier space, was found to possess an one-dimensional 
emergent gauge-like symmetry with a diagonal stripe like structure within 
its ground state.
d) General two spin Fourier kernels, such as 
$v(\vec{k}) = (k^{2}- q^{2})^{2}$ in Eq.(\ref{reduce}),
with a $(d-1)$ dimensional shell of minimizing modes in 
Fourier space are relevant amongst other things to the Brazovskii models 
\cite{Braz}
for liquid crystals, block copolymers, frustrated 
electronic systems, and structural glass transitions \cite{more,glass1}.
We may rigorously establish, in the large $n$ (spherical) limit,
that the ground state entropy
scales as the surface area of
 the system, $S \propto (qN)^{(d-1)/d}$ \cite{explain_n}.
The ground
 state possess an $SO(N^{(d-1)/d})$
 symmetry of which a gauge-like $[U(1)]^\frac{N^{(d-1)/d}}{2}$ is a subgroup.
The Brazovskii system is endowed with a continuous symmetry
harboring a one dimensional gauge-like character.  
Similar to our corollary for gapped quantum systems, in 
the large $n$ Brazovskii system the on-site magnetization is
zero and, in fact, the system possesses a zero temperature
critical point. Indeed in all translationally 
invariant systems with a kernel $v(\vec{k})$ analytic about its
minima and in which the minimizing Fourier modes
form a $(d-1)$ dimensional manifold (as in rotationally invariant 
incommensurate
systems) or a $(d-2)$ dimensional manifold corresponding
to continuous one or two-dimensional 
gauge symmetries. In large $n$ spin systems, any permutation of
the values of the kernel $v(\vec{k}) \to v(P \vec{k})$
(with $P$ any permutation of the $N$ momenta) leaves the partition function
identically invariant \cite{permutation}. This allows an exact reduction
to decoupled chains at low temperatures where 
gauge-like symmetries trivially exist. Such symmetries 
were also shown to rigorously exist in large $n$ renditions of
vectorial spins when subjected to uniform non-Abelian
gauge backgrounds. \cite{glass1} 
[Within the spherical model, the inverse critical
temperature is given by 
\begin{eqnarray}
\frac{1}{k_{B}T_{c}} = \int \frac{d^{d}k}{(2 \pi)^{d}}
\frac{1}{v(\vec{k}) - v(\vec{q})}
\end{eqnarray}
which diverges
for $(d-1)$ or $(d-2)$ dimensional manifolds of modes $\vec{q}$
which minimize an analytic kernel $v(\vec{k})$.] These symmetries 
lead to "zero-energy
 domain walls" which go hand in hand with
fractionalization in the quantum models of \cite{BT}. 
These high degeneracies do not generically survive for finite $n$ system
(where, for example, XY spin systems with the same kernel 
have an emergent discrete $Z_{M}$ symmetry (with $M \sim 
(qL)^{d-1}$)). Nevertheless, 
in some special instances they do, e.g.
\cite{zn01}. 
 
\section{Discussion}

The degeneracy found in all (or only ground) states in systems
having exact (or, respectively, only emergent) gauge-like symmetries
suggests deep non-trivialities 
in the ordering transition
that occurs, if any, at low temperatures.

As well known, in the presence of ground state degeneracies,
entropic fluctuations often drive the system to order
in the vicinity of ``softer'' states having more phase 
space for low energy fluctuations about them \cite{moessner}. 
This mechanism is often termed ``Order out of Disorder''.
The origin of this name are the entropic fluctuations 
(``disorder'') which drive the low temperature ordering transition.
The presence of gauge-like symmetries makes many standard 
methods of analysis much harder or inexact.
Nevertheless, finite temperature
orbital ordering (via an ``order out of 
disorder'' mechanism) was rigorously established
in \cite{NBCv}, \cite{BCN} for two classical systems having
gauge-like symmetries. Finite temperature magnetization 
was proved in the 120 degree model
which harbors an exact $[Z_{2}]^{3L}$ symmetry.
Nematic order 
was also shown to rigorously exist in the classical three dimensional 
orbital compass model which has an exact $[Z_{2}]^{3L}$ symmetry and 
an emergent $[U(1)]^{3L}$ symmetry. The techniques presented 
in \cite{NBCv}, \cite{BCN} may be straightforwardly applied to a host 
of other systems possessing gauge-like symmetries, e.g. 
the effective low energy Hamiltonians of \cite{assa03}, \cite{assa04}
for the pyrochlore and kagome magnets.
The natural
candidates for order parameters are quantities
invariant under all low dimensional gauge-like 
symmetries (exact or emergent). With this guiding principle in 
hand, a natural order parameter is found for 
two-dimensional quantum spin systems \cite{BT} exhibiting 
finite temperature fractionalization.\cite{prep}

In summary, we showed that the generalization of
Elitzur's theorem  to  $d$-dimensional gauge groups
(intermediate between local and global gauge symmetries)
provides a formal meaning for the notion of {\it
dimensional reduction}. Our theorem
proves that for quantities which are non-invariant
under the $d$-dimensional gauge transformations, the
effective dimension of $D$-dimensional theory under
consideration is  reduced from $D$ to $d$.  Elitzur's
theorem is then a consequence of the absence of phase
transitions at any finite temperature in a
zero-dimensional system (see our corollary I).
Moreover, this result also implies that a
one-dimensional gauge symmetry cannot be broken
if the interactions of the theory  have a finite
range (corollary II). The same holds true for continuous
two-dimensional gauge symmetries (corollary III). We
further described different theories of current
interest which possess exact gauge-like 
symmetries for which corollaries II and III
directly apply. Additional 
systems are known to have emergent gauge like
symmetries and we illustrated (corollary IV)
how Elitzur's theorem extends even 
to such systems in the presence of
gaps. The wealth of systems
of current interest for which our theorem 
and its four corollaries directly apply 
illustrates the power and generality of
our result.

This work was sponsored by the US DOE under contracts
W-7405-ENG-36, PICT 03-06343 of ANPCyT and LDRD X1WX.
ZN gratefully acknowledges E. Fradkin
and a productive brief formal visit to UIUC 
which further sparked \cite{NBCF}.


\begin{references}
\bibitem{discrete} For readers wishing to have a simple concrete example 
in mind for which the physical underpinning
can be made transparent, we now sketch matters 
for the two-dimensional orbital compass 
model. Here, the Hamiltonian 
is $H= J \sum_{i} (S_{i}^{x} S^{x}_{i+ \hat{e}_{x}} 
+ S^{y}_{i} S^{y}_{i+ \hat{e}_{y}})$. Insofar as the energetics of spins
on any row (${\cal{C}}_{l}$) parallel to the $x$ axis 
is concerned, the $y$ components
of the spins on nearest neighbor rows act as external magnetic
fields along the $S_{y}$ axis leading to the effective
Hamiltonian, $H = J \sum_{i} S_{i}^{x} S^{x}_{i+ \hat{e}_{x}} 
+ B_{i} S_{i}^{y}$, with $B_{i} = J (S^{y}_{i - \hat{e}_{y}} + 
S^{y}_{i + \hat{e}_{y}})$. Here the fields $\{\eta\}$ of the main 
text are spin components along the row ${\cal{C}}_{l}$
and $\{\psi\}$ includes spins on all other rows
in the system. For a certain choice of the 
spins on all other rows ${\overline{\psi}}$,
we find the Hamiltonian which maximizes the expectation
value $\langle S^{x}_{i} \rangle$ as read off from
the square brackets in 
Eq.(\ref{mast}), $H(\overline{\psi}, \eta) \equiv {\bar H}(\eta)$. 
This corresponds to 
certain external fields $\{\overline{B}_{i}$\} at
the various lattice sites $i$. We now discern the viability of 
finite temperature discrete $Z_{2}$ symmetry 
breaking ($S_{x} \to - S_{x}$) in this system. As well appreciated,
in the absence of an external magnetic field along the 
y-axis ($\overline{B}_{i}=0$), the one dimensional (Ising) system would 
not break $Z_{2}$ symmetry as a consequence of entropic effects 
overwhelming energetic domain wall penalties at all
finite temperatures. This mismatch becomes even more
acute in the presence of finite fields $\{\overline{B}_{i}\}$ 
along the y-axis as the spins become more canted
away from the x-axis, wherein the domain wall
energies are even lower than in a simple one dimensional
Ising model. This reduction in the energy penalty associated with 
symmetry breaking events vis a vis the corresponding energy penalties 
in low dimensional systems- which already would not order on their
own right- is the physical engine behind our 
corollaries.
\bibitem{Bog} To prove this explicitly, 
we follow the procedure of Mermin and 
Wagner and choose the quantities appearing in the 
Bogoliubov inequality appropriately.  As an example,
let us consider the Hamiltonian 
${\bar H} = \frac{1}{2} \sum J_{ij} (S_{i}^{x} S^{x}_{j} 
+ S^{i}_{y} S^{j}_{y}) - \sum_{i} \overline{B}_{i} S_{i}^{z} - 
h S_{\vec{q}}^{x}$
where the infinitesimal external field couples to a given Fourier
component ($\vec{q}$) of $S^{x}$.
For $J_{ij} = J \delta_{|i-j|,1}$, this Hamiltonian depicts
an xy plane in the three dimensional 
orbital compass model (Eq.(\ref{3Dcompass})),
(with a specific $\overline{B}_{i} \equiv  (S_{i+ \hat{e}_{z}}^{z} 
+ S_{i- \hat{e}_{z}}^{z})$
maximizing the square brackets in Eq.(\ref{mast})). 
We now set, in the Bogoliubov inequality 
$|\langle [C^{\dagger}, A^{\dagger}]\rangle |^{2}
\le \frac{1}{2T} (AA^{\dagger} + A^{\dagger}A) 
\langle [C^{\dagger},[{\bar H},C]] 
\rangle$, the quantities to be $C= S_{\vec{k}}^{z}$ (the Fourier component 
conjugate to the 
external $\overline{\psi}$) and $A=S^{x}_{-\vec{k}-\vec{q}}$
(corresponding to $\eta$). In the thermodynamic limit (wherein 
the number of lattices sites $N \to \infty$), this leads to a 
vanishing $m_{\vec{q}} = N^{-1} \langle S_{\vec{q}}^{x} \rangle$ 
at all finite temperatures 
when the external field applied  $h \to 0$ in a system of short range 
interactions such that $\frac{1}{2N} \sum_{ij} J_{ij} 
|\vec{x}_{i} - \vec{x}_{j}|^{2}$ (where $\vec{x}_{i}$ is the spatial location
of site $i$) is bounded. 
Here, the external ``magnetic fields'' $\overline{B}_{i}$ acting along the 
z-axis do not spoil the invariance of the Hamiltonian
under the symmetry group $G_{j}$ (corresponding to $O(2)$
rotations in the xy plane).
\bibitem{explain_replica} The central idea of
the replica formulation is that
the free energy averaged over in
the quenched expectation value
is trivially given by the identity 
$\ln Z = \lim_{n \to 0} \frac{Z^{n} - 1}{n}$ 
with $Z$ the partition function and $n$ the 
number of replicas.
\bibitem{explain_n} The general ground state is given by
the real space spin $S(\vec{r}_{i}) = \sum_{m} a_{q_{m}} \cos(\vec{q}_{m}
\cdot \vec{r}_{i} + \phi_{m})$ (with $\{\vec{q}_{m}\}$
the set of modes minimizing the interaction kernel $v(\vec{k})$)
subject to the spherical normalization 
constraint $\sum_{m} [a_{q_{m}}]^{2} = N^{2}$.
There are $(qL)^{d-1}$ amplitudes subject to 
a single normalization constraint and thus
the ground state entropy scales as 
$(qL)^{d-1}$.
\bibitem{permutation}
In the notation of Eq.(\ref{reduce})
for spin polarization independent spectra
$v_{\alpha \beta} (\vec{k}) = v(\vec{k}) \delta_{\alpha \beta}$, 
the partition function of the spherical (large $n$)
model is 
\begin{eqnarray}
Z = \prod_{\vec{k}}  [\beta(v(\vec{k}) + \mu)]^{-1/2},
\label{Zperm}
\end{eqnarray}
with $\mu$ determined from 
\begin{eqnarray}
\beta = \int \frac{d^{d}k}{(2 \pi)^{d}} [v(\vec{k}) + \mu]^{-1}
\label{chem}
\end{eqnarray}
with $\beta$ the inverse temperature.
As is evident from Eqs.(\ref{Zperm},\ref{chem}),
for any of the $N!$ permutations $P$ of the spectra
 $\{v(\vec{k})\}
 \to \{v(P\vec{k})\}$,
the partition function $Z$ is invariant.
\end{references}
\end{document}